\documentclass[runningheads]{llncs}
\usepackage{graphicx}
\usepackage{siunitx}
\usepackage[square,numbers]{natbib}
\usepackage{tabularx}
\usepackage[titletoc,title]{appendix}
\usepackage{amsmath}
\begin{document}
\title{Retinal Vessel Segmentation based on Fully Convolutional Networks}
%
%
\author{Zhengyuan Liu}
\authorrunning{Z. Liu} 
\titlerunning{UCLA CS M.S. Capstone Project Report} 
%
\institute{University of California, Los Angeles \\
\email{zyliu@cs.ucla.edu}}
\maketitle              
\begin{abstract}
The morphological attributes of retinal vessels, such as length, width, tortuosity and branching pattern and angles, play an important role in diagnosis, screening, treatment, and evaluation of various cardiovascular and ophthalmologic diseases such as diabetes, hypertension and arteriosclerosis. The crucial step before extracting these morphological characteristics of retinal vessels from retinal fundus images is vessel segmentation. In this work, we propose a method for retinal vessel segmentation based on fully convolutional networks. Thousands of patches are extracted from each retinal image and then fed into the network, and data argumentation is applied by rotating extracted patches. Two architectures of fully convolutional networks, U-Net and LadderNet, are used for vessel segmentation. The performance of our method is evaluated on three public datasets: DRIVE, STARE, and CHASE\_DB1. Experimental results of our method show superior performance compared to recent state-of-the-art methods.

\end{abstract}
\section{Introduction}

The morphological attributes of retinal vessels, such as length, width, tortuosity and branching pattern and angles, play an important role in diagnosis, screening, treatment, and evaluation of various cardiovascular and ophthalmologic diseases such as diabetes, hypertension and arteriosclerosis~\cite{fraz2012blood}. The crucial step before extracting these morphological characteristics of retinal vessels from retinal fundus images is vessel segmentation. However, manually segmenting vessels in retinal images is both error-prone and time-consuming, even for experienced physicians. Therefore, it is vital to develop accurate and automatic methods to segment blood vessels from retinal images.

Image segmentation, with the goal to assign semantic labels (vessel and background in the case of retina vessel segmentation) to every pixel in an image, is one of the fundamental topics in computer vision, especially in biomedical image processing~\cite{chen2018encoder,ronneberger2015u}. Deep convolutional neural networks can automatically learn an increasingly complex hierarchy of features directly from the input data, and show striking improvement over systems relying on hand-crafted features on many visual recognition tasks~\cite{chen2018encoder}, including biomedical image segmentation~\cite{hatamizadeh2019deep,hatamizadeh2019lesion,hatamizadeh2019end,imran2018automatic}. Specifically, Ronneberger et al.~\cite{ronneberger2015u} proposed U-Net architecture with a contracting path and a symmetric expansive path based on fully convolutional network~\cite{long2015fully} and used data augmentation by applying elastic deformations to the limited training biomedical images, which outperformed prior state-of-the-art models in several different biomedical segmentation challenges.

Deep convolutional neural networks have also been applied to retinal vessel segmentation tasks in recent work. Melin\v{s}\v{c}ak et al.~\cite{melinvsvcak2015retinal} addressed vessel segmentation using a 10-layer CNN. Liskowski and Krawiec~\cite{liskowski2016segmenting} used a structured prediction scheme to highlight context information, and tested a comprehensive set of architectures among which a 7-layer no-pooling CNN was the most successful. Fu et al.~\cite{fu2016deepvessel} combined a typical 7-layer CNN with a conditional random field, reformulated as a recurrent neural network, to model long-range pixel interactions. Oliveira et al.~\cite{oliveira2018retinal} proposed a novel architecture composed of an encoder and a decoder based on fully convolutional network~\cite{long2015fully}, while using rotation operations for data augmentation and feeding extra channels into the network through the SWT decomposition. Recently, a couple of advanced network architectures, including M2U-Net~\cite{laibacher2018m2u}, R2U-Net~\cite{alom2018recurrent} and LadderNet~\cite{zhuang2018laddernet}, have been applied to retinal vessel segmentation tasks, and achieved superior performance in experimental results.

In this work, we propose a method for retinal vessel segmentation based on fully convolutional networks. Thousands of patches are extracted from each retina image and then fed into the network, and data argumentation is applied by rotating extracted patches. Two architectures of fully convolutional networks, U-Net~\cite{ronneberger2015u} and LadderNet~\cite{zhuang2018laddernet}, are used for vessel segmentation. The performance of our method is evaluated on three public datasets: DRIVE~\cite{niemeijer2004comparative}, STARE~\cite{hoover2000locating}, and CHASE\_DB1~\cite{owen2009measuring}. We use accuracy and area under the ROC curve as major evaluation metrics. Experimental results of our method show superior performance compared to recent state-of-the-art methods.

The remaining sections are organized as follows. Section 2 describes our method and the fully convolutional networks used for retinal vessel segmentation. Experiments setup is described in Section 3. Our results are shown in Section 4. Finally, we come to the conclusion in Section 5. 

\section{Method}
The goal of retinal vessel segmentation is to assign one of the two labels, i.e. vessel (1) and background (0), to each pixel of the retinal image, thus it is a binary classification problem. There are three main stages in our proposed method for retinal vessel segmentation: preprocessing on the retinal image, patch extraction and data augmentation, and classification via fully convolutional networks. In this Section, we first describe the datasets used for retinal vessel segmentation, and then illustrate the three stages of our method in detail below.

\subsection{Dataset}
\begin{figure}
\includegraphics[width=\textwidth]{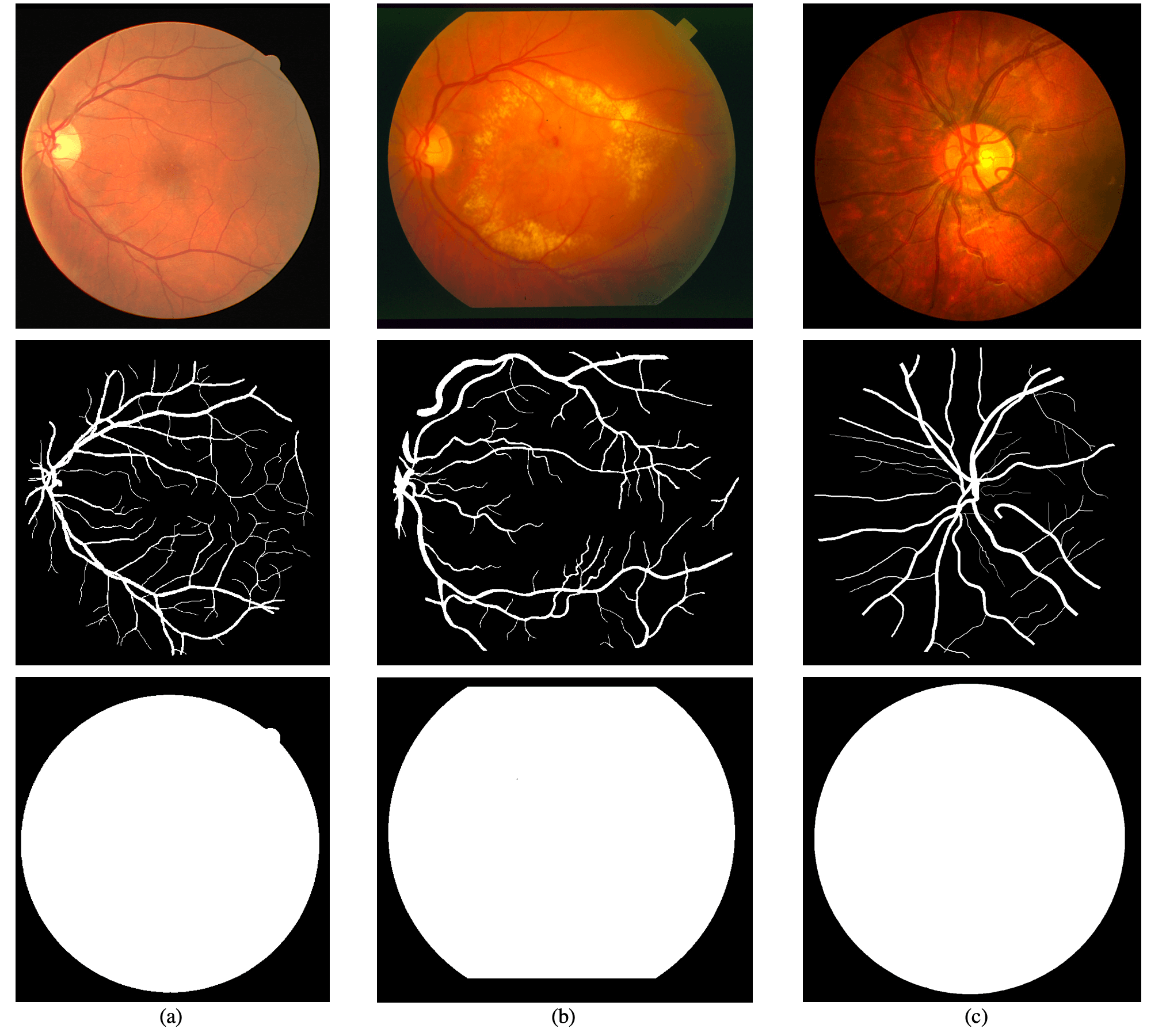}
\caption{Sample images with their ground truth vessel segmentation masks and FOV masks from (a) DRIVE, (b) STARE and (c) CHASE\_DB1, respectively.} 
\label{fig_data}
\end{figure}

3 public retinal vessel segmentation datasets are used to train and evaluate the proposed method: DRIVE~\cite{niemeijer2004comparative}, STARE~\cite{hoover2000locating}, and CHASE\_DB1~\cite{owen2009measuring}. Sample images from three datasets with their ground truth vessel segmentation masks and field of view (FOV) masks are shown in Fig.~\ref{fig_data}. Pixels outside the FOV are not imaged and inessential for the vessel segmentation, so we only care about pixels inside the FOV. FOV masks for DRIVE images are originally available in the dataset, while STARE and CHASE\_DB1 do not originally contain FOV masks. Thus we downloaded FOV masks for STARE and CHASE\_DB1 from the supporting GitHub repository of Oliveira et al.~\cite{oliveira2018retinal}\footnote{\texttt{https://github.com/americofmoliveira/VesselSegmentation\_ESWA}}. There are two sets of manual segmentations available in all three datasets, and we use the first annotations as ground truth for all of them.

The DRIVE dataset consists of 40 images, and each image has a resolution of $565 \times 584$. The STARE dataset provides 20 images for vessel segmentation tasks, where 10 images are diagnosed as normal while other 10 images are diagnosed as pathological. Each image has a resolution of $700 \times 605$. The CHASE\_DB1 dataset has 28 images, collected from two eyes of 14 children. Each image has a resolution of $999 \times 960$.

In the DRIVE dataset, the set of 40 images has already been divided into a training set and a test set, both containing 20 images, so the test set is directly used to evaluate our method. However, there is no division of training and test sets for STARE and CHASE\_DB1. Therefore, we used the stratified $k$-fold cross-validation method to evaluate our method, where the original dataset was partitioned into $k$ equal-sized folds, and each fold is used as the test set while images in the remaining $k - 1$ folds are used for training. The evaluation process, including training and testing the model, was repeated $k$ times for each fold used as testing set. The results were then averaged to get the final evaluation result of the model. For the STARE dataset, $k$ is set to 5, where each fold has 4 images and 2 of them are normal while 2 of them are pathological. For the CHASE\_DB1 dataset, $k$ is set to 4, where each fold has 7 images and 3 images are left/right eyes while 4 images are the other.

\subsection{Preprocessing}
\begin{figure}
\includegraphics[width=\textwidth]{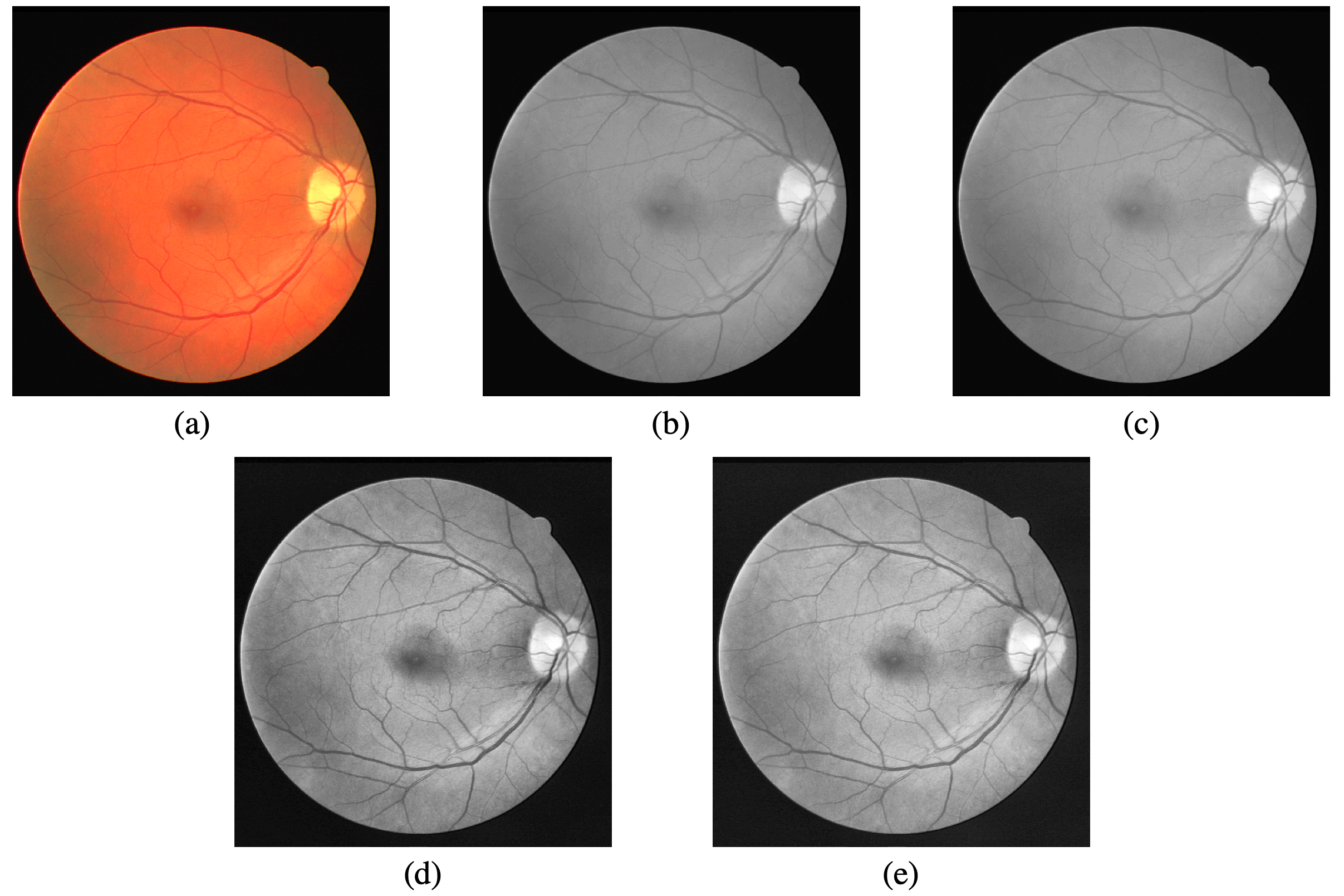}
\caption{Preprocessing results after each step: (a) Original retina image. (b) Gray-scale conversion. (c) Standardization. (d) CLAHE. (e) Gamma adjustment.} 
\label{fig_pre}
\end{figure}

The first stage of our method is preprocessing on the retinal images. There are four steps in the preprocessing of the original images: gray-scale conversion, Standardization, Contrast-limited adaptive histogram equalization (CLAHE) and Gamma adjustment. The result images after each step are shown in Fig.~\ref{fig_pre}.

\subsubsection{Gray-scale Conversion}
Retina images are typically 3-channel RGB images, as shown in Fig.~\ref{fig_pre} (a). By converting these images into gray-scale images (Fig.~\ref{fig_pre} (b)), it eases subsequent preprocessing steps and reduces computational overhead in training. For each pixel, the formula used to convert images from RGB to gray-scale is:
$I = 0.299\times R + 0.587\times G + 0.114 \times B$
where $R$, $G$, $B$ denote the pixel's values of R, G, B channels respectively and $I$ denotes the pixel's new value, i.e. intensity, in the gray-scale image.

\subsubsection{Standardization}
Z-score standardization is applied to every pixel of the gray-scale images:
$$x_i' = \frac{x_i - \overline{X}}{\sigma_X}$$
where $x_i$ and $x_i'$ denote the pixel's original and new value, while $\overline{X}$ and $\sigma_X$ denote the mean and standard deviation of pixel values from all of the gray-scale images. By applying Z-score standardization, the distribution of the pixel values is centralized, which will speed up the training phase and improve generalization~\cite{han2011data}.

We subsequently applied a min-max normalization to map all the pixel values to range [0.0, 1.0], since for images with float value-type, their pixel values should in range [0.0, 1.0] by convention. The result image of standardization is shown in Fig.~\ref{fig_pre} (c).

\subsubsection{CLAHE}
Adaptive histogram equalization (AHE) is a contrast enhancement method for both natural and medical images, which computes several histograms corresponding to distinct sections of the image and uses them to redistribute the lightness values of the image. However, AHE tends to over-amplify the contrast in near-constant regions of the image and over-enhances noise. Contrast Limited AHE (CLAHE) reduces the noise amplification problem by applying limitation on contrast amplification~\cite{pizer1987adaptive}. The result image of applying CLAHE is shown in Fig.~\ref{fig_pre} (d).

\subsubsection{Gamma adjustment}
Gamma adjustment is an exponential transform operating on pixel’s intensity that usually used for contrast adjustment:
$$J = 255(\frac{I}{255}) ^ {1/\gamma}$$
where $I$ denotes pixel’s original intensity and $J$ denotes pixel’s intensity after adjustment. When $\gamma > 1$, contrast in low-intensity region is enhanced and contrast in high-intensity region is decreased; when $\gamma < 1$, contrast in low-intensity region is decreased and contrast in high-intensity region is enhanced. Here $\gamma$ is set to larger than 1, in order to enhance the contrast in low-intensity region so that the vessels in dark regions will be more distinct. The result of gamma adjustment is shown in Fig.~\ref{fig_pre} (e).

\begin{figure}
\includegraphics[width=\textwidth]{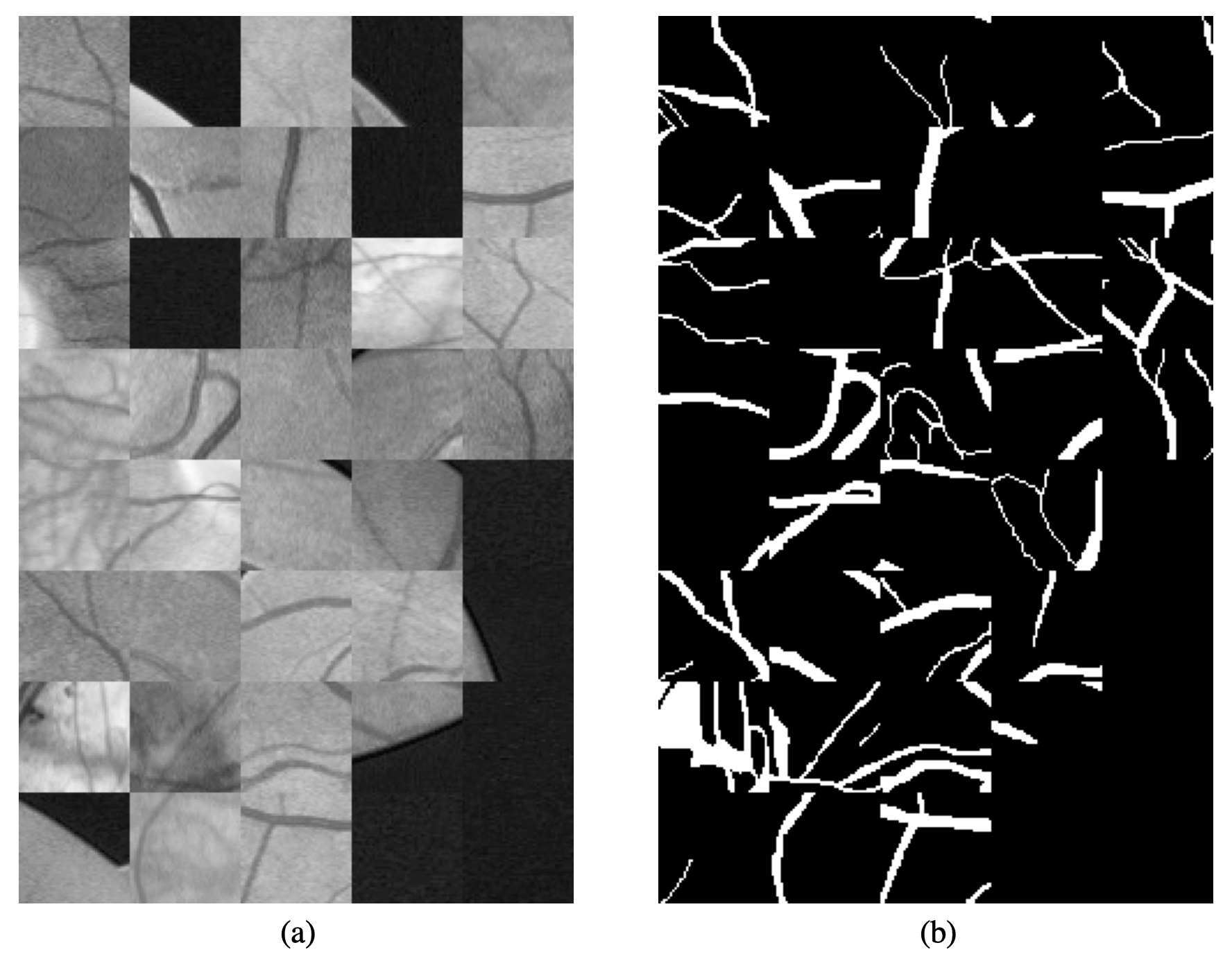}
\caption{Sample inputs of the network: (a) input patches from preprocessed images, and (b) their corresponding ground truth segmentation mask patches.} 
\label{fig_input}
\end{figure}

\subsection{Patch Extraction and Data Augmentation}
Since the number of images in each retinal vessel segmentation dataset is very limited (40 in DRIVE, 20 in STARE and 28 in CHASE\_DB1), we extracted a set of patches from preprocessed images, and fed these patches, instead of full images, into the network for training. During testing, images are also cut into patches for inference. 

During training, 2000 patches with size $48 \times 48$ are randomly extracted from each image. All the pixels in the patch are guaranteed inside the image, but not guaranteed inside field of view. Although patches may overlap with each other, this process tremendously enlarged the training dataset and made training of the fully convolutional network feasible and efficient.

During testing, images are cut into patches with size $48 \times 48$ and fed into the network for inference. The most straightforward way to cut images into patches is cutting seamlessly, i.e. there is no overlap between different patches. However, in this case some pixels that are originally inner pixels of the full image become borders of patches which will bring bias when combining segmentation maps of the patches into the segmentation mask for the full image. Specifically, in some case the output segmentation mask looks like it consists of patches, rather than a naturally full image segmentation. This problem will be described in more details in Section 4. In order to solve this problem and improve the performance, we applied another way that cut the images into patches by using a stride of $N$ pixels (e.g. $N = 5$ or $10$) in both height and width, so that one pixel can be predicted for multiply times. The final inference result is obtained by averaging multiple predictions. In addition, zero padding are applied to the original image to making its size become an integral multiple of the patch size in both ways.
Besides patch extraction, we also applied data augmentation by rotating the patches, since data augmentation is essential to teach the network the desired invariance and robustness properties~\cite{ronneberger2015u}. Each extracted patch was rotated by \ang{90}, \ang{180} and \ang{270}, thus the training set is quadrupled, i.e. 8000 patches are extracted from each image after data augmentation. At the end, all the patches, including original extracted and generated by rotation, are randomly shuffled and then fed into the network together with their corresponding ground truth segmentation mask patches. Sample inputs of the network are shown in Fig.~\ref{fig_input}.

\subsection{Network Architecture}
Two architectures of fully convolutional networks, U-Net and LadderNet, are used for vessel segmentation. The architecture of these two networks are described in details below.

\subsubsection{U-Net}
\begin{figure}
\includegraphics[width=\textwidth]{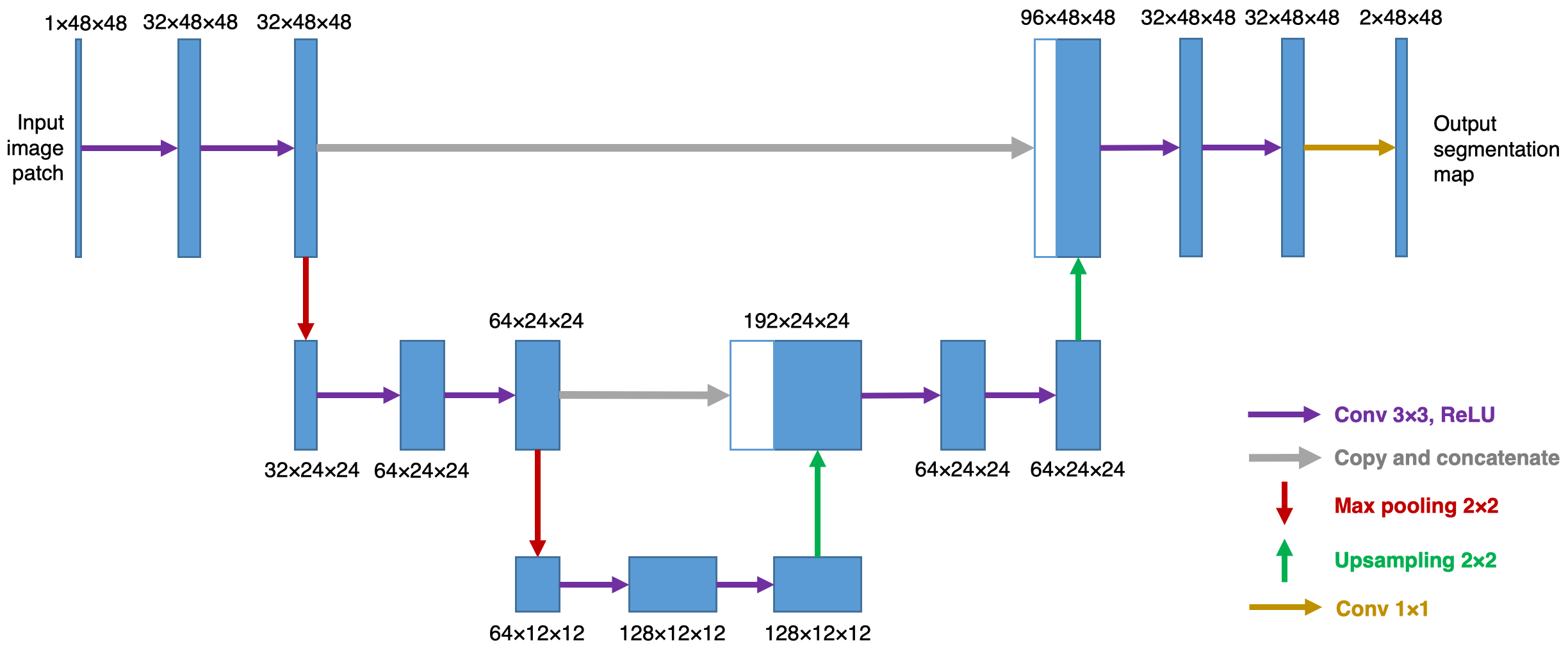}
\caption{The architecture of U-Net. Each blue box refers to a multi-channel feature map in the network, and the shape of each feature map (depth-x-y) is shown on top of the box. White boxes refer to copied feature maps. Different colors of arrows denote different kinds of operations.} 
\label{fig_unet}
\end{figure}

Our U-Net architecture is inspired by U-Net~\cite{ronneberger2015u} and the fully convolutional network proposed by Oliveira et al.~\cite{oliveira2018retinal}, which combines features from both architectures. The network architecture is shown in Fig.~\ref{fig_unet}. The network is composed of an encoder (i.e. the contracting path at left side) and a decoder (i.e. the symmetric expansive path at right side).

The main idea of this encoder-decoder architecture is to combine information with different levels of abstraction. In the encoder, several pooling operators successively reduce the dimensions of the feature maps produced by each convolutional layer. This encoder structure summarizes neighboring features and creates high-level information maps, however, with low resolution. While these maps contain rich and compact features and are suitable for recognizing objects in images, detailed information related to localization is lost. Therefore, in the decoder, deep, low-resolution, high-level feature maps are reconstructed and combined with shallow, high-resolution, low-level information layers from the encoder through skip connections. The high-resolution, low-level information available in the shallow layers can help to recover the spatial information and improve localization performance. This is the main motivation for the decoder structure. 

In our network, the input layer is a 1-channel patch extracted from preprocessed retinal images as described in previous sections, then follows a contracting path, i.e. encoder, and an expansive path, i.e. decoder. Every step in the contracting path consists of two consecutive $3 \times 3$ convolutions (with zero padding to remain the same size), followed by a $2 \times 2$ max pooling operation with stride 2 for downsampling. After each max pooling in the contracting path, the number of feature channels is doubled, from 32, to 64, and finally 128. Every step in the expansive path consists of a $2 \times 2$ upsampling of the feature map, followed by two consecutive $3 \times 3$ convolutions (with zero padding) and a concatenation with the corresponding feature map from the contracting path through skip connections. After each upsampling in the expanding path, the number of feature channels is halved. Therefore, the expansive path is symmetric to the contracting path, and yields a U-shaped architecture like U-Net~\cite{ronneberger2015u}. Finally, a $1 \times 1$ convolution is used to output the segmentation maps (with 2 channels for 2 classes in this case) from the last 32-channel feature maps. Rectifier Linear Unit (ReLU) is used as the activation function after each convolutional layer, and a dropout rate of 0.2 is used between two consecutive convolutional layers to helps prevent overfitting~\cite{srivastava2014dropout}.

\subsubsection{LadderNet}
\begin{figure}
\includegraphics[width=\textwidth]{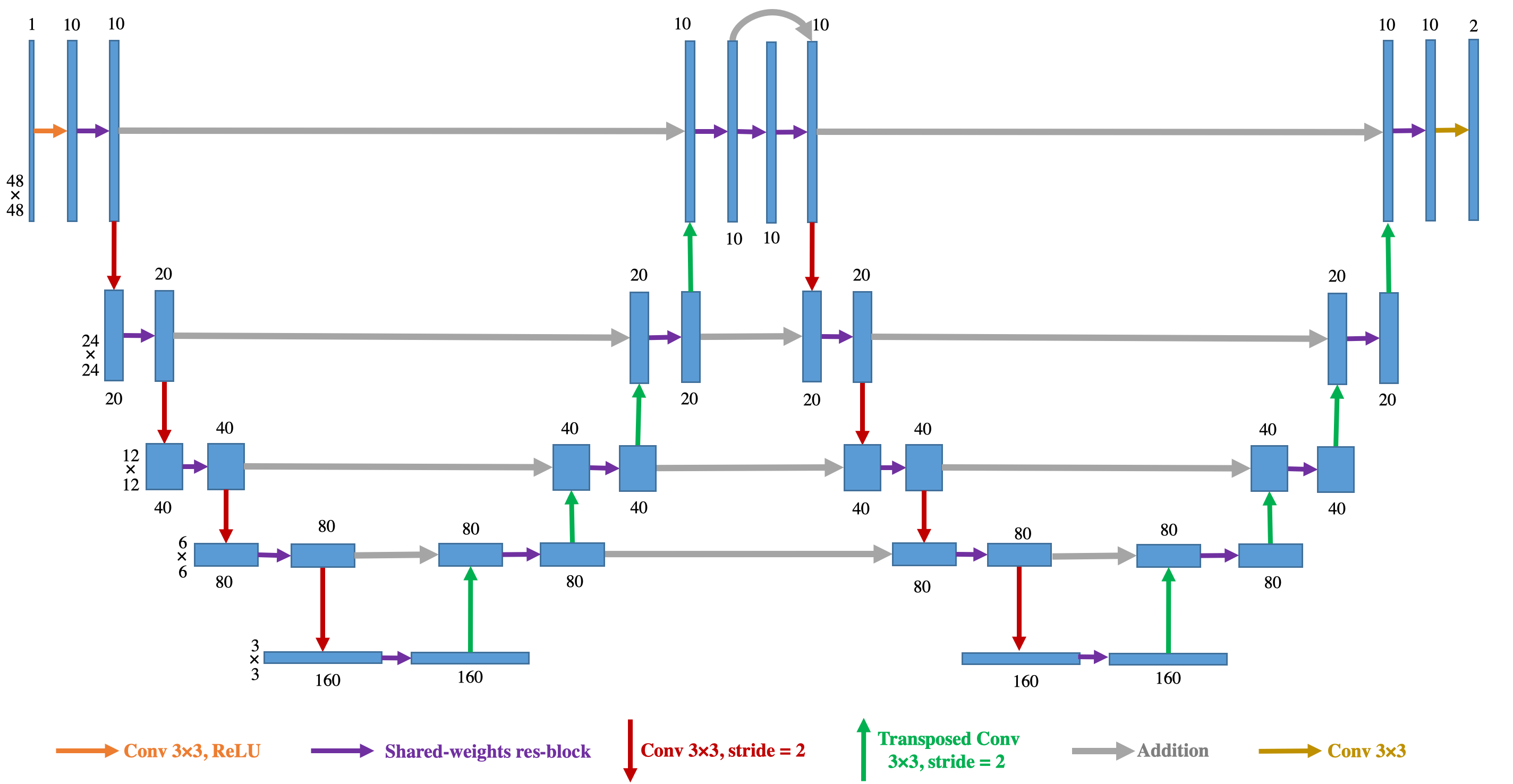}
\caption{The architecture of LadderNet. Each blue box refers to a multi-channel feature map in the network. The number of channels is shown on top of the box. The size (width-height) is shown at left of each layer. Different colors of arrows denote different kinds of operations.} 
\label{fig_laddernet}
\end{figure}

LadderNet~\cite{zhuang2018laddernet} is a multi-branch fully convolutional network (FCN), which can be viewed as a chain of U-Nets. Instead of only one pair of encoder-decoder branch in U-Net, LadderNet has multiple pairs of encoder-decoder branches, and has skip connections between each pair of adjacent decoder and decoder branches in each level. The architecture of LadderNet is shown in Fig.~\ref{fig_laddernet}. Different from U-Net where feature maps in the encoder branch are concatenated with feature maps in the decoder branches, feature maps are summed from two branches.

LadderNet can also be viewed as an ensemble of multiple FCNs. LadderNet provides multiple paths of information flow, not only across the encoder-decoder branch in one U-Net, but also across multiple encoder-decoder branches from different U-Nets. Each path can be viewed as a variant of FCNs. LadderNet has much more information flow than U-Net, therefore has the potential to capture more complicated features.

\begin{figure}
\includegraphics[width=\textwidth]{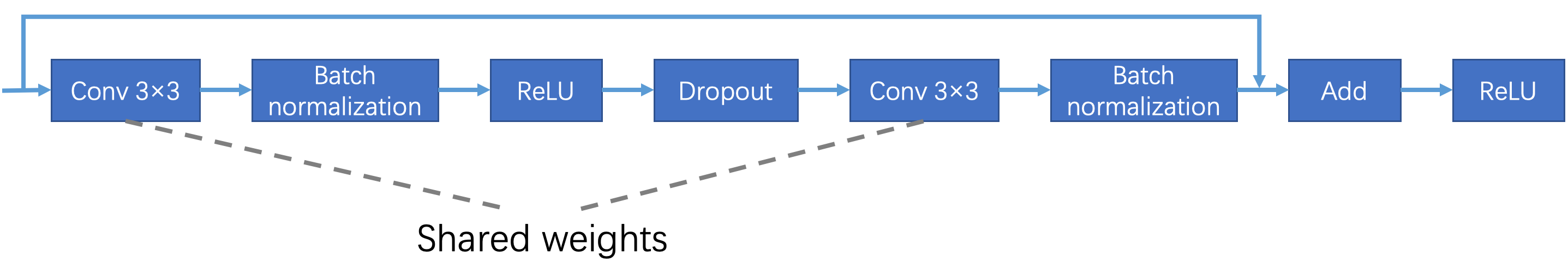}
\caption{The detailed structure of a shared-weights residual block.} 
\label{fig_res-block}
\end{figure}

However, multiply U-Nets and encoder-decoder branches will increase the number of parameters and the difficulty of training. To solve this problem, To solve this problem, shared-weights residual blocks are used in LadderNet (Fig.~\ref{fig_res-block}). Different from a standard residual convolutional block proposed by He~\cite{he2016deep}, the two convolutional layers in the same block share the same weights. A dropout layer is added between two convolutional layers to avoid overfitting. The shared-weights residual block combines the strength of skip connection, recurrent convolution, batch normalization and dropout, and has much fewer parameters that a standard residual block.

\section{Experiments}
\subsection{Experimental Setup}
Training and testing of the proposed network was implemented by Keras\footnote{\texttt{https://keras.io/}} with TensorFlow\footnote{\texttt{https://www.tensorflow.org/}} backend. The experiments are conducted on a MacBook Pro 2016 with a 2.7 GHz Intel Core i7 CPU and 16GB RAM, and running macOS Mojave 10.14.3. Because of the limitation of computation resource (this machine does not support GPU acceleration for TensorFlow), we downloaded a pretrained U-Net model from a GitHub repository\footnote{\texttt{https://github.com/orobix/retina-unet}}, which has already been trained for 150 epochs on DRIVE dataset, and then kept training all three datasets on this pretrained model. For LadderNet, we downloaded a pretrained model from the supporting GitHub repository of Zhuang~\cite{zhuang2018laddernet}\footnote{\texttt{https://github.com/juntang-zhuang/LadderNet}}. 

\subsection{Training}
As described before, 2000 patches are randomly extracted from each image in the dataset and then rotated by \ang{90}, \ang{180} and \ang{270} for data augmentation. For DRIVE dataset, 160000 patches are obtained for training from the 20 images in the training set. For STARE dataset, 128000 patches are obtained for training, as the original image set was divided into a 16-image training set and a 4-image test set. For CHASE\_DB1 dataset, there are 168000 patches obtained for training, as the original image set was divided into a 21-image training set and a 7-image test set. In all experiments, 90\% of these extracted patches are used for training while the remaining 10\% are used for validation. All three datasets are training on the pretrained U-Net model for 10 epochs, with a mini-batch size of 32 patches. For LadderNet, DRIVE dataset is training on the pretrained model for 4 epochs, with a mini-batch size of 1024 patches.

\subsection{Evaluation Metrics}
We used several most commonly used metrics in the literature to evaluate the performance of our method: accuracy (Acc), sensitivity (Sn), specificity (Sp), and F1-score (F1). The formulas to compute these metrics are shown below:
$$Acc = \frac{TP+TN}{TP+TN+FP+FN}$$
$$Sn = \frac{TP}{TP+FN}$$
$$Sp = \frac{TN}{TN+FP}$$
$$Precision = \frac{TP}{TP+FP}$$
$$Recall = \frac{TP}{TP+FN}$$
$$F1 = 2 \times \frac{Precision \times Recall}{Precision+Recall}$$
We further calculated the precision-recall curve and receiver operating characteristics (ROC) curve, and used the area under the ROC curve (AUC) as our final metric.

\section{Discussion and Results}
\subsection{Discussion}
\begin{figure}
\includegraphics[width=\textwidth]{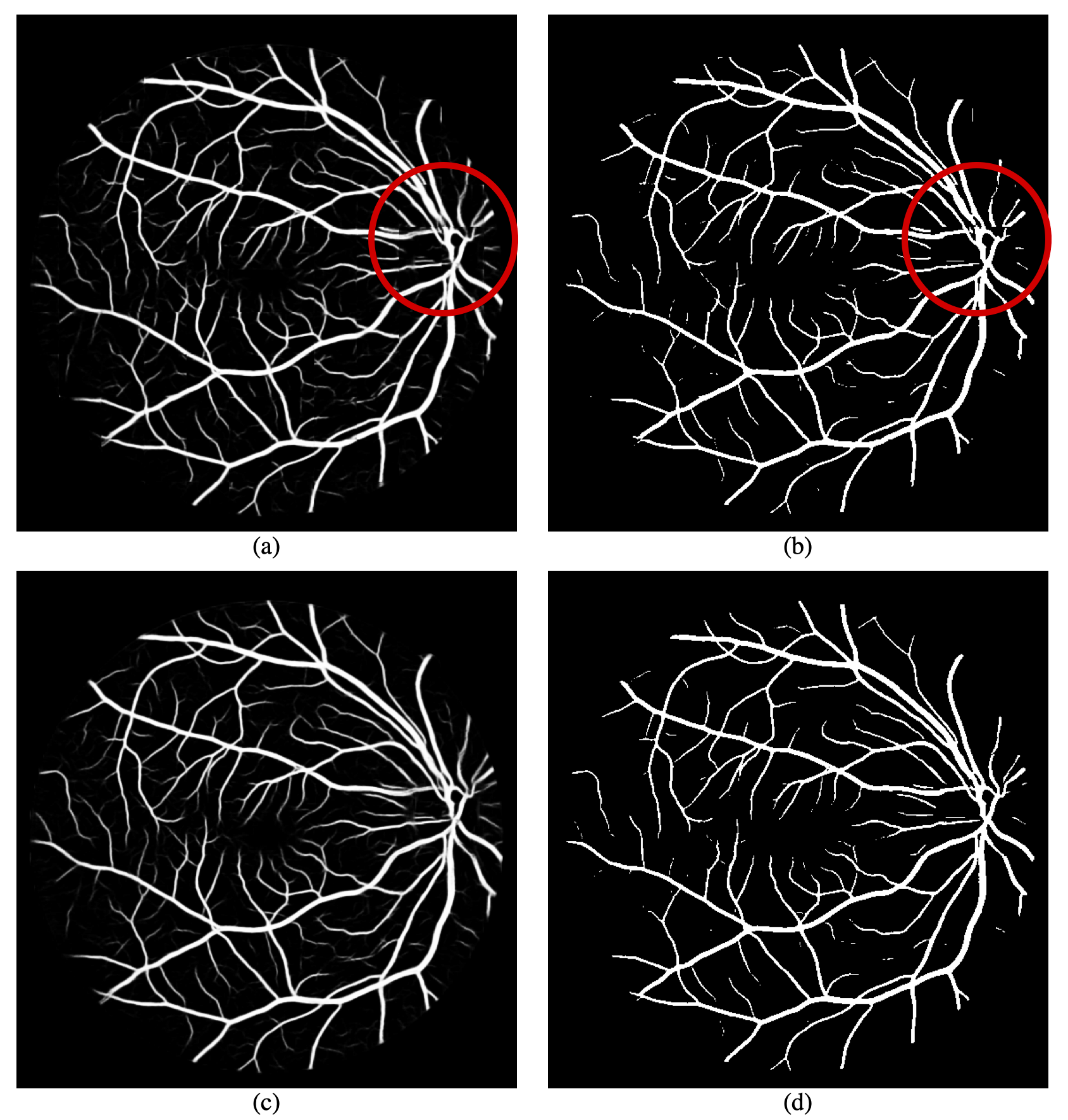}
\caption{An example of the issue of using Non-overlap patches in inference: (a) probability map of Non-overlap; (b) segmentation mask of Non-overlap; (c) probability map of Stride = 5; (d) segmentation mask of Stride = 5. The red circles marked the unnatural parts in the output of Non-overlap. Using a stride of 5 alleviate this issue.} 
\label{fig_discuss}
\end{figure}

In this section, we discussed how different methods of cutting images into patches in the inference stage, i.e. cutting images into seamless and non-overlap patches (Non-overlap) and cutting images into overlapping patches by using a stride (Stride), would affect the performance of segmentation. As describe in Section 2.3, in the first way, some pixels that are originally inner pixels of the full image become borders of patches, which will bring bias when combining segmentation maps of the patches into the segmentation mask for the full image. Specifically, in some case the output segmentation mask looks like it consists of patches, rather than a naturally full image segmentation. One example of this issue is shown in Fig.~\ref{fig_discuss} (a) and (b). In order to alleviate this problem, we cut the images into overlapping patches by using a stride of $N$ pixels (e.g. $N = 5$ or $10$) in both height and width, and final inference results are obtained by averaging multiple predictions. In this way, one pixel can be predicted for multiply times, which can amortize the border effect and output better segmentation results (Fig.~\ref{fig_discuss} (c) and (d)). 

Besides generate more natural segmentation results for the full image, overlapping patches by using a stride $N$ can further improve the performance of the segmentation and achieve a higher Acc. AUC and F1-score (as shown in Table~\ref{tab_discuss}). When using a smaller $N$, the performance will keep improved, however the inference time for one image will also increase. In the subsequent experiments, we used a stride of 5. In the real-world application, however, we can set a larger stride to reduce inference time if needed.

\begin{table}
\centering
\caption{Results and inference time for one image of different inference methods on DRIVE dataset.}\label{tab_discuss}
\setlength{\tabcolsep}{1.5mm}
\begin{tabular}{c|c|c|c|c|c|c}
\hline
Cutting Method & \multicolumn{6}{c}{DRIVE}\\
\cline{2-7} & Time(s) & F1 & Sn & Sp & Acc & AUC\\
\hline
Non-overlap & 3.66 & 0.8117 & 0.7701 & 0.9814 & 0.9545 & 0.9760\\
Stride = 20 & 22.80 & 0.8156 & 0.7731 & 0.9821 & 0.9555 & 0.9785\\
Stride = 10 & 68.96 & 0.8163 & 0.7723 & 0.9825 & 0.9557 & 0.9790\\
Stride = 5 & 257.37 & 0.8169 & 0.7728 & 0.9826 & 0.9559 & 0.9794\\
\hline
\end{tabular}
\end{table}

\subsection{Results}
\begin{figure}
\includegraphics[width=\textwidth]{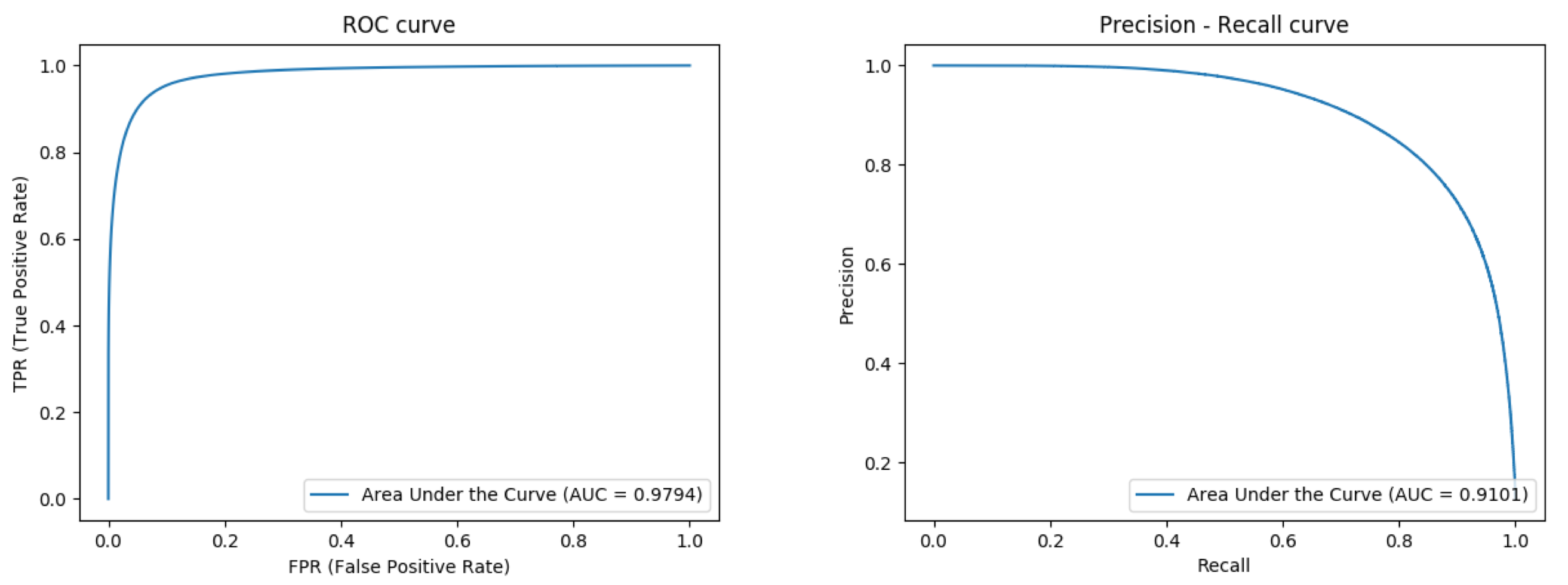}
\caption{ROC curve and Precision-Recall curve on DRIVE datasets.} 
\label{fig_drive_roc}
\end{figure}

\begin{figure}
\includegraphics[width=\textwidth]{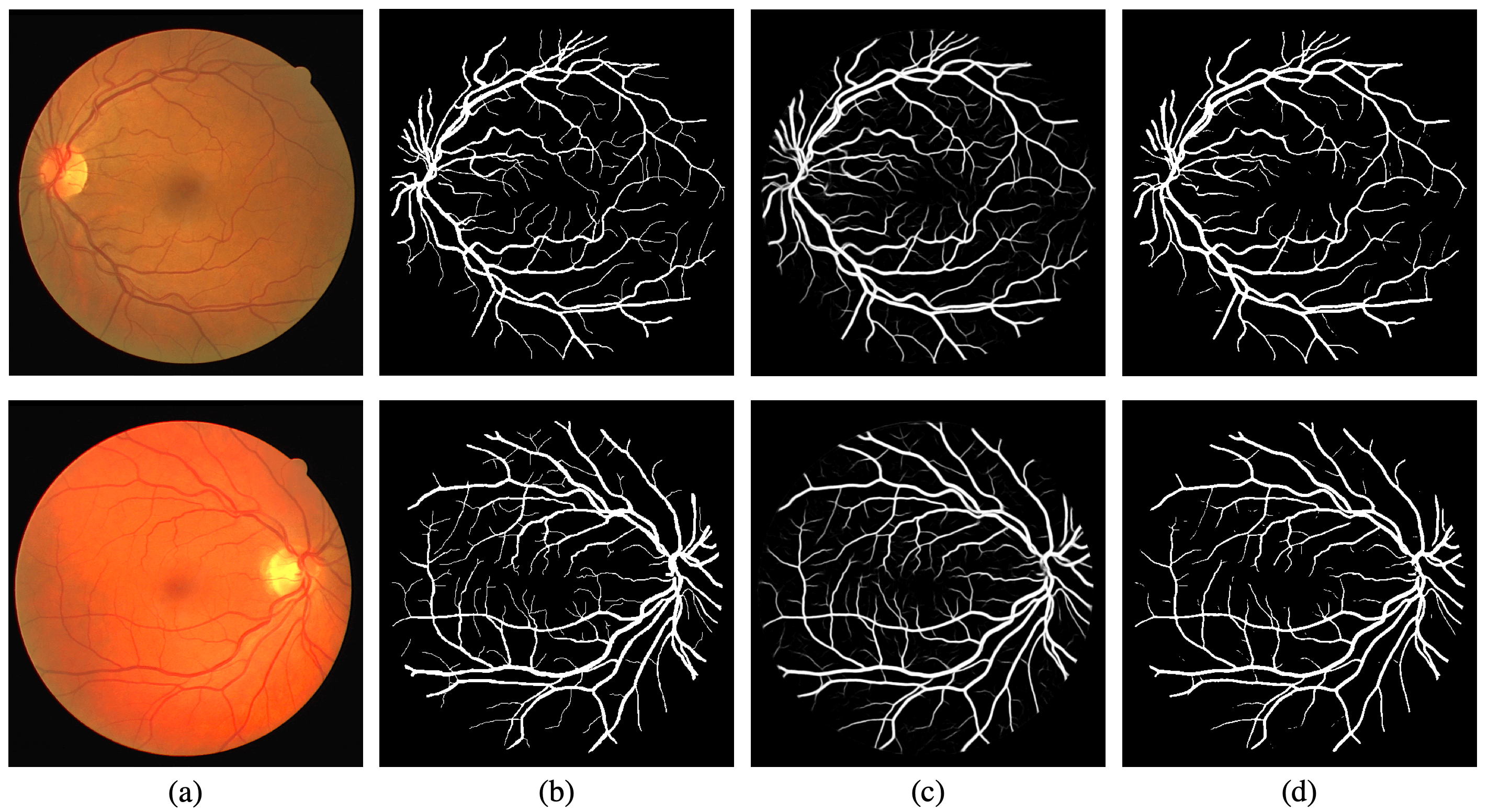}
\caption{Segmentation results for DRIVE dataset: (a) retinal image; (b) ground truth; (c) output probability map; (d) binary segmentation.} 
\label{fig_drive_res}
\end{figure}

\begin{figure}
\includegraphics[width=\textwidth]{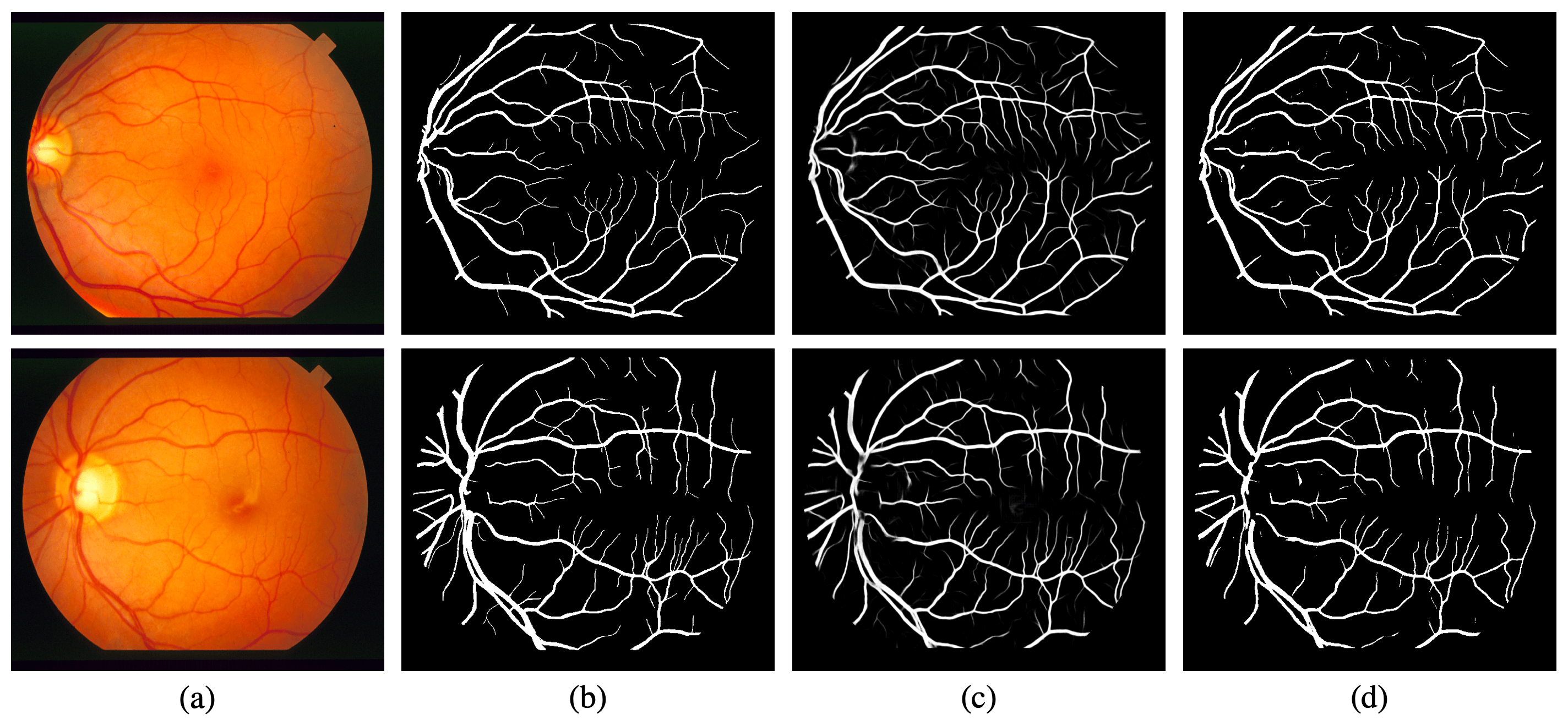}
\caption{Segmentation results for STARE dataset: (a) retinal image; (b) ground truth segmentation; (c) output probability map; (d) binary segmentation.} 
\label{fig_stare_res}
\end{figure}

\begin{figure}
\includegraphics[width=\textwidth]{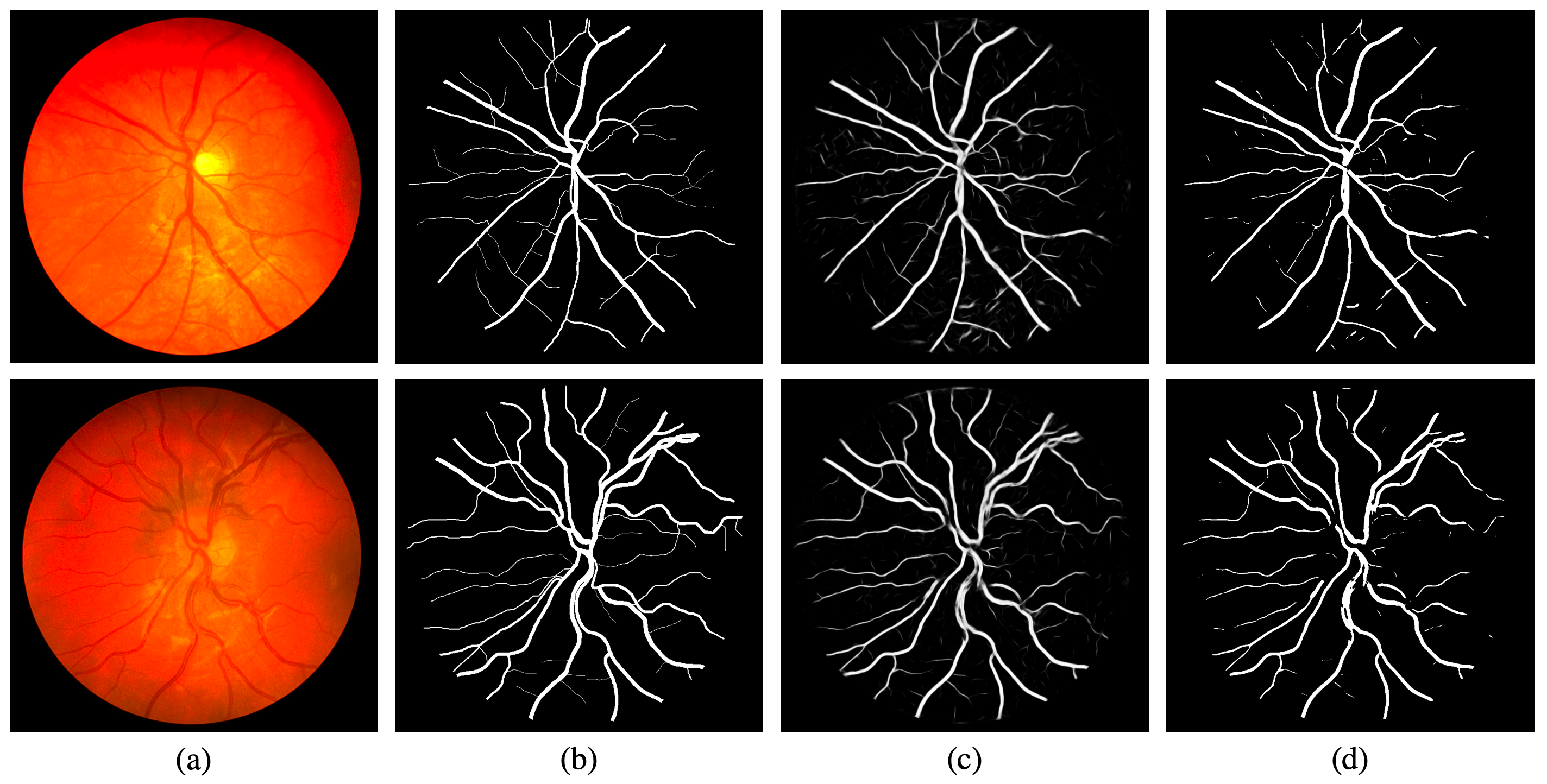}
\caption{Segmentation results for CHASE\_DB1 dataset: (a) retinal image; (b) ground truth segmentation; (c) output probability map; (d) binary segmentation.} 
\label{fig_chase_res}
\end{figure}

Because of the limitation of space, we only report our results of U-Net in detail, and briefly report our results of LadderNet. The segmentation results (including output probability map and binary segmentation) of our U-Net for images in the test set of DRIVE are shown in Fig.~\ref{fig_drive_roc} and~\ref{fig_drive_res}. Samples of segmentation results of our U-Net for STARE and CHASE\_DB1 are shown in Fig.~\ref{fig_stare_res} and~\ref{fig_chase_res} respectively. The threshold used to generate the binary segmentation is 0.5.

The performance of our method and other state-of-the-art deep learning based methods on DRIVE, STARE and CHASE\_DB1 are shown in Table~\ref{tab_res_drive},~\ref{tab_res_stare} and~\ref{tab_res_chase} respectively. The results show that our method has superior performance compared to recent state-of-the-art methods.

\begin{table}
\centering
\caption{Segmentation results of different deep learning based methods on DRIVE. Bold values show the best score among all methods.}\label{tab_res_drive}
\setlength{\tabcolsep}{1mm}
\begin{tabular}{c|c|c|c|c|c}
\hline
Method & \multicolumn{5}{c}{DRIVE}\\
\cline{2-6} & F1 & Sn & Sp & Acc & AUC\\
\hline
Melin\v{s}\v{c}ak et al.~\cite{melinvsvcak2015retinal} & - & 0.7276 & 0.9785 & 0.9466 & 0.9749\\
Li et al.~\cite{li2015cross} & - & 0.7569 & 0.9816 & 0.9527 & 0.9738\\
Liskowski et al.~\cite{liskowski2016segmenting} & - & 0.7520 & 0.9806 & 0.9515 & 0.9710\\
Fu et al.~\cite{fu2016deepvessel} & - & 0.7603 & - & 0.9523 & -\\
Oliveira et al.~\cite{oliveira2018retinal} & - & \textbf{0.8039} & 0.9804 & 0.9576 & \textbf{0.9821}\\
M2U-Net~\cite{laibacher2018m2u} & 0.8091 & - & - & \textbf{0.9630} & 0.9714\\
R2U-Net~\cite{alom2018recurrent} & 0.8171 & 0.7792 & 0.9813 & 0.9556 & 0.9784\\
LadderNet~\cite{zhuang2018laddernet} & 0.8202 & 0.7856 & 0.9810 & 0.9561 & 0.9793\\
\hline
\textbf{This work, U-Net} & 0.8169 & 0.7728 & \textbf{0.9826} & 0.9559 & 0.9794\\
\textbf{This work, LadderNet} & \textbf{0.8219} & 0.7871 & 0.9813 & 0.9566 & 0.9805\\
\hline
\end{tabular}
\end{table}

\begin{table}
\centering
\caption{Segmentation results of different deep learning based methods on STARE. Bold values show the best score among all methods.}\label{tab_res_stare}
\setlength{\tabcolsep}{1mm}
\begin{tabular}{c|c|c|c|c|c}
\hline
Method & \multicolumn{5}{c}{STARE}\\
\cline{2-6} & F1 & Sn & Sp & Acc & AUC\\
\hline
Li et al.~\cite{li2015cross} & - & 0.7726 & 0.9844 & 0.9628 & 0.9879\\
Liskowski et al.~\cite{liskowski2016segmenting} & - & 0.8145 & 0.9866 & 0.9696 & 0.9880\\
Fu et al.~\cite{fu2016deepvessel} & - & 0.7412 & - & 0.9585 & -\\
Oliveira et al.~\cite{oliveira2018retinal} & - & \textbf{0.8315} & 0.9858 & 0.9694 & 0.9905\\
R2U-Net~\cite{alom2018recurrent} & \textbf{0.8475} & 0.8298 & 0.9862 & \textbf{0.9712} & \textbf{0.9914}\\
\hline
\textbf{This work, U-Net} & 0.8219 & 0.7739 & \textbf{0.9867} & 0.9638 & 0.9846\\
\textbf{This work, LadderNet} & 0.7694 & 0.7513 & 0.9764 & 0.9529 & 0.9660\\
\hline
\end{tabular}
\end{table}

\begin{table}
\centering
\caption{Segmentation results of different deep learning based methods on CHASE\_DB1. Bold values show the best score among all methods.}\label{tab_res_chase}
\setlength{\tabcolsep}{1mm}
\begin{tabular}{c|c|c|c|c|c}
\hline
Method & \multicolumn{5}{c}{CHASE\_DB1}\\
\cline{2-6} & F1 & Sn & Sp & Acc & AUC\\
\hline
Li et al.~\cite{li2015cross} & - & 0.7507 & 0.9793 & 0.9581 & 0.9716\\
Fu et al.~\cite{fu2016deepvessel} & - & 0.7130 & - & 0.9489 & -\\
Oliveira et al.~\cite{oliveira2018retinal} & - & 0.7779 & \textbf{0.9864} & 0.9653 & \textbf{0.9855}\\
M2U-Net~\cite{laibacher2018m2u} & 0.8006 & - & - & \textbf{0.9703} & 0.9666\\
R2U-Net~\cite{alom2018recurrent} & 0.7928 & 0.7756 & 0.9820 & 0.9634 & 0.9815\\
LadderNet~\cite{zhuang2018laddernet} & \textbf{0.8031} & \textbf{0.7978} & 0.9818 & 0.9656 & 0.9839\\
\hline
\textbf{This work, U-Net} & 0.7815 & 0.7240 & 0.9860 & 0.9600 & 0.9784\\
\hline
\end{tabular}
\end{table}

\section{Conclusion}
In this work, we presented a method for retinal vessel segmentation using patch-based fully convolutional networks. Patches are extracted from the retinal image and then fed into the networks, and data argumentation is applied by rotating extracted patches. We applied two network architectures for vessel segmentation: U-Net and LadderNet. Experimental results of our method show superior performance compared to recent state-of-the-art methods.

The limitation of our vessel segmentation method is the robustness and connectivity of the output vessel segmentation, as some thin vessels are in small, broken bits and not connected to the main vessel tree. The borders of the patch also have some side-effects when generating the segmentation mask for the full image. For future work, more advanced network architectures, such as DenseNet~\cite{huang2017densely} or DeepLabv3+~\cite{chen2018encoder}, can be used to further improve the performance and make the segmentation output more robust. Moreover, large-size patches, even full images, can be used as the input to the networks, to give the networks more information from the input and alleviate the size-effects of patches.

\section*{Acknowledgement}
I would like to acknowledge the help from Prof. Demetri Terzopoulos and valuable discussions with Dr. Ali Hatamizadeh from UCLA Computer Graphics \& Vision Laboratory.

%
%
%
%
\bibliographystyle{splncs04}
\bibliography{ref}

\end{document}